\documentclass[showpacs,preprintnumbers,nofootinbib,amsmath,amssymb]{revtex4}

\usepackage{graphicx}

\begin{document}

\title{Axial Stiffness of Multiwalled Carbon Nanotubes}

\author{Vladimir Zavalniuk}
\email[]{VZavalnyuk@onu.edu.ua}%
\affiliation {Department of Theoretical Physics, Odessa II Mechnikov National University, \\2 Dvoryanskaya St., Odessa 65026, Ukraine}

\begin{abstract}
The axial stiffness of MWCNTs is demonstrated to be determined only by several external shells (usually 3-5 and up to 15 for the extremely large nanotubes and high elongations) what is in a good agreement with experimentally observed inverse relation between the radius and Young modulus (i.e., stiffness) of MWCNTs. This result is a consequence of the van der Waals intershell interaction. The interpolating formula is obtained for the actual axial stiffness of MWCNT as a function of the tube external radius and elongation.
\end{abstract}

\pacs{61.48.De,62.25.-g,62.20.de}
%\submitto{Phys. Rev B}

\maketitle
% ----------------------------------------------------------------
\subsection{Introduction}

The unusual and even sometimes wonderful mechanical properties of carbon nanotubes and their bundles make it possible to use them just now for the wide range of applications. As an example, nanotubes can act as an reinforcement of different materials (plastics, hydrocarbon resins, nanocomposites etc.) where their extremal bending flexibility and axial stiffness are of a great interest.\cite{Reinf1,Reinf2,Reinf3}

It is established that SWCNTs can sustain strains larger than 10\% of tensile deformation prior to fracture \cite{fracture1,fracture2} and their deformation is completely reversible (i.e., elastic) subjected to strains of more than 4\%.\cite{Revers1,Revers2,Revers3,Revers4} A lot of works were dedicated to investigation of elastic properties of single-walled nanotubes (SWCNT) and multiwalled nanotubes (MWCNT). \cite{MWCNTDepOnRadius1,fracture2,Revers3,Revers4,Young1,Young2,Young3,Young4,Young5,Young6MW,Young8thCont,Young9} Theoretically MWCNTs were studied for uniform axial stresses at both of their ends,\cite{Young7thMW} but this is not the only possible loading type. Apparently in most cases of axial tension only the external shell of MWCNT is affected by the imposed load and internal shells are involved into the considered process only due to van der Waals intershell interaction (the simplest case of such situation is the deformation of the capped MWCNT).

It is obvious that in the case of uniformly deformed $n$-walled CNT the intershell distances $\Delta r_{i}=r_{i}-r_{i+1}=(1-\nu\varepsilon)d_{0}$ remain equal (where $i=1,2,\dots,n\!-\!1$ and $i=1$ corresponds to the outermost shell), but when the load is imposed only on the external shell the distances between other shells are not equal and increase with $i$ (for $\varepsilon>0$). Due to strong nonlinear dependence of the intershell interaction on the intershell distance the difference in deformation energy between both mentioned cases for certain specific elongation may be significant and should be studied.

Some experiments showed that the effective Young modulus of the MWCNT is inversely proportional to its radius (for radii in range from 4 nm to 20 nm) \cite{InverseRadiusPoncharal} and such results contradict the assumption of the equal deformations of shells. Some theoretical works also showed the dependense of the effective Young modulus on the number of wall \cite{MWCNTDepOnRadius1,MWCNTDepOnRadius2} while other did not show any dependence on radius even for SWCNTs.\cite{SWCNTNotDepOnRadius,Young7thMW}

In this paper we use the stiffness $k\sim Y d$ instead of Young modulus $Y$ in order to avoid the uncertain parameter "wall thickness" $d$. Some authors stand on using this parameter because when considering flexural deformations of nanotube (within the string approximation) we should work with $Y d^3$ but it is evident that we always can switch from one pair of independent parameters to another such pair (for example, from $Y$ and $d$ to $k$ and $\gamma$, where $\gamma$ denotes some flexural characteristic).

\subsection{Axial Stiffness of SWCNT}

Since the investigation of stiffness (or Young modulus) of single-walled nanotubes is not the main objective of this work we will describe it briefly presenting here without derivation only most important for our purposes expressions.

In the simplest case the axial stiffness of a cylindric shell (such as a nanotube) with surface density $m_{0} \tau$ ($m_{0}$ is the mass of each atom and $\tau$ is their quantity per unit area) is defined by the expression
\begin{equation}\label{LongitudinalRigidity}
k(R,L) = 2\pi m_{0} \tau c^{2}\frac{R}{L} = \alpha \frac{R}{L},
\end{equation}
where $R$ and $L$ are the shell radius and length, $c$ is the longitudinal sound velocity and $\alpha$ is the constant parameter for graphene-based structures. Certainly previous expression is correct for nanotubes of different radii only if shells properties are not radius-dependent. In the case of nanotubes with extremely small radii ($R \sim d_{0}$ where $d_{0}\!=\!0.34$ nm is distance between graphitic planes conditioned by the van der Waals interaction) the re-hybridization of atomic orbitals lead to perceptible changes of its mechanical properties (the raising of effective Young modulus is observed on quantum dynamics simulations for SWCNTs with $R\!<\!6$ nm \cite{MolecDyn1,MolecDyn2}), but for tubes with radii more than one nanometer such changes may be ignored.

The sound velocity $c$ may be obtained from the microscopic models of nanotube (or graphene) as a velocity of acoustic phonons. Further we will use the value $c\!\!=18.4$ km/sec \cite{Adamyan} which lead to $\alpha\!\approx\!1632$ \,kg sec$^{-2}$. If necessary the expressions for effective Young modulus and Poisson ratio also may be derived by comparing equations of motion from both microscopic and continuum models.

By (\ref{LongitudinalRigidity}) one can calculate the \textit{idealized} Young modulus of SWCNT
\[
Y=k\frac{L}{S}=\alpha\frac{R}{S}=\alpha\frac{R}{2\pi R d} \approx 0.73 \,T Pa,
\]
where $S$ is the effective surface of the nanotube cross section and the commonly accepted value of the "wall thickness" $d=d_{0}$ is used.
On the other hand the experimentally measurable \textit{effective} Young modulus of the macroscopic bundle of SWCNTs with equal radii $R$ depend not on the effective surface of its cross section but on the total cross section surface of the bundle $S_{b} \approx N \pi (R+d_{0}/2)^{2}$ (where $N$ is the number of tubes in the bundle):
\[
Y_{b}=N k \frac{L}{S_{b}}=N \alpha\frac{R}{S_{b}}=\alpha \left.\frac{R}{\pi (R+d_0/2)^2} \right|_{R\gg d_{0}} \approx \frac{\alpha}{\pi} \, \frac{1}{R}.
\]
One can see that the effective Young modulus of such a bundle should be inversely proportional to the nanotubes average radius.

\subsection{Van der Waals intershell interaction}

\begin{figure}[!hbp]
\includegraphics[scale=1.5]{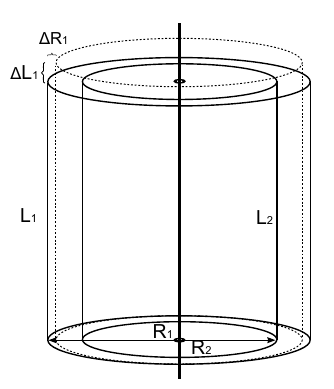}
\caption{The definition of the main parameters of the considered DWCNT. }\label{Cylinders}
\end{figure}

Within the continuum approximation the van der Waals intershell interaction energy for the DWCNT depends only on shells radii ($r_1, r_2$) and lengthes ($L_1, L_2$) [Fig. \ref{Cylinders}]. In terms of hypergeometric functions it can be expressed as follows \cite{Marchenko} (assuming that $L_1 \geq L_2$)
\begin{equation}\label{eqInteractionEnergyIdeal}
U_{0}(R_{1},L_{1},R_{2},L_{2})=\frac{3}{2} \pi^{3} \tau^{2} R_{1} R_{2} L_{2}
\left( \frac{21}{32} \, \gamma_{12} \frac{\Phi\left(\frac{11}{2},R_{1},R_{2}\right)}{(R_{1}+R_{2})^{11}} - \gamma_{6} \frac{\Phi\left(\frac{5}{2},R_{1},R_{2}\right)}{(R_{1}+R_{2})^{5}} \right),
\end{equation}
where $\gamma_{6}=2.43\times10^{-24} \,\, \textrm{J}{\cdot}\textrm{nm}^{6}$ and $\gamma_{12}=3.859\times10^{-27}\,\, \textrm{J}\cdot\textrm{nm}^{12}$ are attractive and repulsive constants of the Lennard-Jones potential,\cite{girifalco} $\tau$ is the surface density of carbon atoms and
\[
\Phi\left(J,R_{1},R_{2}\right) := \,_{2}\textrm{F}_{1} \left( \frac{1}{2}, J, 1, \frac{4R_{1}R_{2}}{(R_{1}+R_{2})^{2}} \right) =
\frac{(R_{1}+R_{2})^{2J}}{2\pi} \int_{-\pi}^{\pi}\frac{d \theta}{(R_{1}^{2}+R_{2}^{2}-2R_{1}R_{2}cos\theta)^{J}}.
\]
In the case of unstrained graphene (or nanotube) $\tau\!=\!\tau_{0}\!=\!\frac{4}{3 \sqrt{3} \,\, b^{2}}\!=\!38.2$ nm$^{-2}$ ($b\!=\!0.142$ nm).

The length and radius of the nanotube under the tension take the values $l\!=\!L_{0}(1+\varepsilon)$ and $r\!=\!R_{0}(1-\nu\varepsilon)$ where $\varepsilon$ is the specific elongation of nanotube, $\nu=0.17$ \cite{PoissonRatio1970,PoissonRatioT,PoissonRatioG} is the graphene Poisson ratio, $L_{0}$ and $R_{0}$ are length and radius of unstrained nanotube. In this case the nanotube surface $S$ and the surface density of atoms change correspondingly:
$$S_{j}=2\pi R_{j} L_{j} (1-\nu\varepsilon_{j})(1+\varepsilon_{j}),
\qquad \tau_{j}=\frac{\tau_{0}}{(1+\varepsilon_{j})(1-\nu\varepsilon_{j})}.$$

The interaction energy (\ref{eqInteractionEnergyIdeal}) for the case of strained shells ($\varepsilon_{1}$ and $\varepsilon_{2}$ respectively) takes on form
\begin{equation}\label{eqInteractionEnergyStrained}
\begin{array}{c}
  U(R_{1},L_{1},\varepsilon_{1},R_{2},L_{2},\varepsilon_{2}) = \frac{3}{2} \pi^{3} \tau_{1}\tau_{2} r_{1} r_{2} l_{2}
\left( \frac{21}{32} \, \gamma_{12} \frac{\Phi(\frac{11}{2},r_{1},r_{2})}{(r_{1}+r_{2})^{11}} - \gamma_{6} \frac{\Phi\left(\frac{5}{2},r_{1},r_{2}\right)}{(r_{1}+r_{2})^{5}} \right) = \\
 = \frac{3}{2} \pi^{3} \frac{1}{1+\varepsilon_{1}} \tau^{2} R_{1} R_{2} L_{2}
\left(
\frac{21}{32} \, \gamma_{12}
 \frac{\Phi\left(\frac{11}{2},R_{1}(1-\nu\varepsilon_{1}),R_{2}(1-\nu\varepsilon_{2})\right)} {\left(R_{1}(1-\nu\varepsilon_{1})+R_{2}(1-\nu\varepsilon_{2})\right)^{11}}
- \gamma_{6}
 \frac{\Phi\left(\frac{5}{2},R_{1}(1-\nu\varepsilon_{1}),R_{2}(1-\nu\varepsilon_{2})\right)} {\left(R_{1}(1-\nu\varepsilon_{1})+R_{2}(1-\nu\varepsilon_{2})\right)^{5}}
\right).
\end{array}
\end{equation}

From now on we will denote by $U_{W}$ the \textit{van der Waals deformation} energy that is the contribution of the intershell interaction into the total deformation energy of double-walled nanotube
\begin{equation}\label{eqVdWDeformationEnergy}
\Delta U_{W}(R_{1},L_{1},\varepsilon_{1},R_{2},L_{2},\varepsilon_{2}):=U(R_{1},L_{1},\varepsilon_{1},R_{2},L_{2},\varepsilon_{2}) - U_{0}(R_{1},L_{1},R_{2},L_{2}).
\end{equation}

\subsection{Axial Stiffness of DWCNT}

The total deformation energy of DWCNT consists from the deformation energies of both shells and the additional (to unstrained state) intershell interaction energy:
\begin{equation}\label{TotalDefEnergyOfDWCNT1}
\begin{array}{c}
E(\varepsilon_{1},\varepsilon_{2},R_{1},R_{2},L_{1},L_{2}) =
\frac{k_{1} \Delta L_{1}^2}{2} + \frac{k_{2} \Delta L_{2}^2}{2} + \Delta U_{W}(R_{1},L_{1},\varepsilon_{1},R_{2},L_{2},\varepsilon_{2}) =\\
\frac{k_{1} L_{1}^{2} \varepsilon_{1}^2}{2} + \frac{k_{2} L_{2}^{2} \varepsilon_{2}^2}{2} + \Delta U_{W}(R_{1},L_{1},\varepsilon_{1},R_{2},L_{2},\varepsilon_{2})=\\
\frac{\alpha}{2}\left(R_1 L_1 \varepsilon_{1}^2 + R_2 L_2 \varepsilon_{2}^2\right) + \Delta U_{W}(R_{1},L_{1},\varepsilon_{1},R_{2},L_{2},\varepsilon_{2}).
\end{array}
\end{equation}

If the both shells are of equal length ($L_1\!=\!L_2\!=\!L$), the total deformation energy of DWCNT will be linear in its length and the inner shell relative deformation will depend only on that of the outer shell and shells radii.

We assume that the external force acts only on the outer shell and as a result the DWCNT axial stiffness should be introduced as
$$k_{DW}(\varepsilon_1)=\left.\frac{1}{L_1^2}\frac{\partial^2 E}{\partial \varepsilon_1^2}\right|_{E=\min E(...,\varepsilon_{2},...)}$$

\begin{figure}[!hbp]
\includegraphics[scale=1.0]{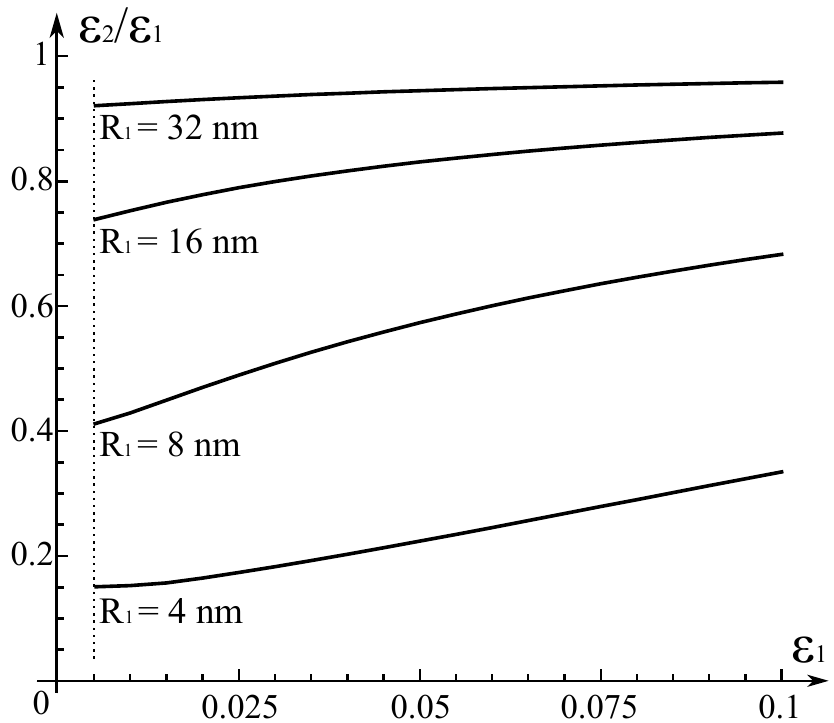}
\caption{The dependence between the outer and inner shell relative deformations ($\varepsilon_1$ and $\varepsilon_2$ correspondingly) for DWCNTs of different radii and unstrained intershell distance $d_0$. For very small deformations the behavior of a DWCNT is determined by its actual unstrained intershell distance which is not likely to be equal to the energetically optimal distance $d_{0}$ due to the nanotube's radius discrete nature. }\label{Epsilons}
\end{figure}

Obviously, $\varepsilon_2\!<\!\varepsilon_1$ when $R_1 \nu\,\varepsilon_1\!<\!d_0$, and for the fixed value of $\varepsilon_1$ the $\varepsilon_2$ will increase strongly with increasing of DWCNT radius [Fig. \ref{Epsilons}].

\subsection{Axial Stiffness of MWCNT}

Like that of DWCNT the total deformation energy of $N$-walled MWCNT is determined by the following expression
\begin{equation}\label{TotalDefEnergyOfMWCNT}
E(\epsilon,\mathbf{R},\mathbf{L},N) =
\frac{\alpha}{2}\sum\limits_{i=1}^{N}R_{i} L_{i} \varepsilon_{i}^{2} +
\sum\limits_{i=1}^{N-1}\Delta U_{W}(R_{i},L_{i},\varepsilon_{i},R_{i+1},L_{i+1},\varepsilon_{i+1}),
\end{equation}
where $\epsilon=(\varepsilon_{1},...,\varepsilon_{N})$, $\mathbf{R}=(R_{1},...,R_{N})$ and $\mathbf{L}=(L_{1},...,L_{N})$.

The axial stiffness in this case is defined as follows
\begin{equation}\label{RigidityOfMWCNT}
k_{MW}(\varepsilon_1,\mathbf{R},\mathbf{L},N)=\left.\frac{1}{L_1^2}\frac{\partial^2 E(\epsilon_{min},\mathbf{R},\mathbf{L},N)}{\partial \varepsilon_1^2}\right|_{E(\epsilon_{min},...)=\min_{\varepsilon_{2},...,\varepsilon_{N}} E(\epsilon,\mathbf{R},\mathbf{L},N)}.
\end{equation}

\begin{figure}[!hbp]
\includegraphics[scale=1.0]{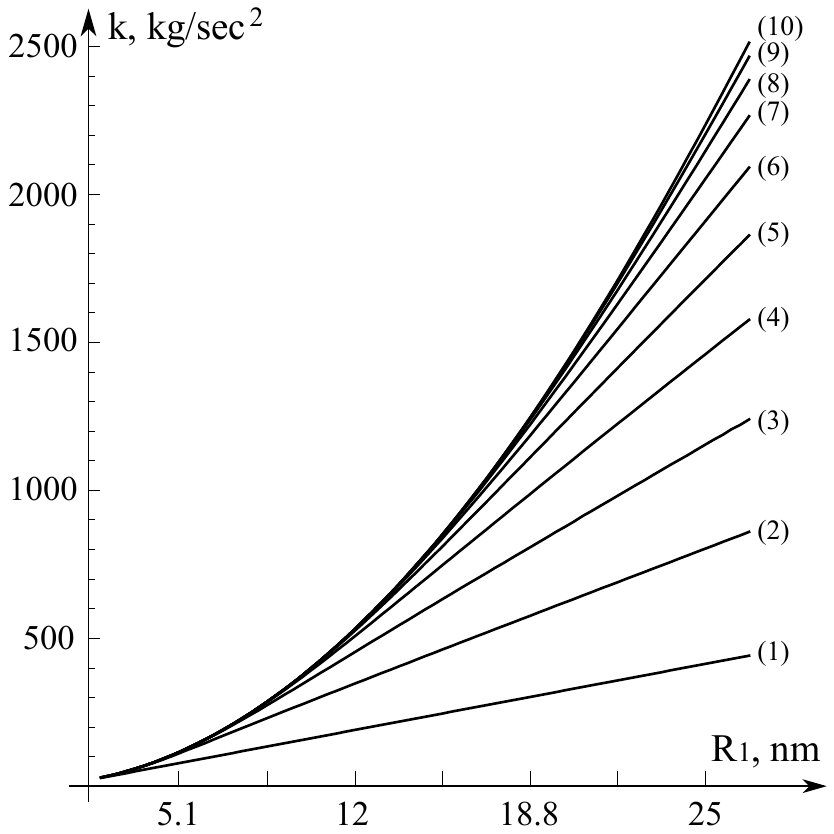}
\caption{The dependence of MWCNT stiffness on the tube's external radius $R_{1}$ and the number of its shells (in brackets) for the $\varepsilon_1=0.05$ and $L=100$ nm. The asymptotic value of the number of shells involved into the deformation is about 15 even for the extremely large MWCNTs (for tubes with $R<10$ nm there only four shells are actually involved).}\label{RigidityMWCNT}
\end{figure}

Unlike the DWCNT all but innermost shells of MWCNT have the following neighbor which hampers the shell's deformation of comparing to the case of DWNCT so that the relative deformation  of inner shells $\varepsilon_{i}$ decrease rapidly with $i$. As a result the actual axial stiffness of MWCNT grows with number of shells much slower than its "ideal" stiffness $k_{ideal}=\sum k(R_{i},L_{i})$. Actually for MWCNTs with external radius $R\leq 10$ nm and arbitrarily number of shells under the strain with $\varepsilon_1\lesssim0.05$ only four external shells contribute to the total stiffness and this number grows slowly up to $\approx 10$ for the nanotubes with $R>25$ nm [Fig. \ref{RigidityMWCNT}].

For $N$-walled MWCNTs with external radii more than $3$ nm (assuming that $N$ is greater than the number of shells actually involved into the deformation) the stiffness can be fitted by the following expression with an accuracy of 1-3 per cents (for $0<\varepsilon_{1}<0.1$)
\begin{equation}\label{MaxRigidityOfMWCNT}
k(R,L,\varepsilon)\!=\!\frac{10^{-7}}{L} \exp\!\left[2.3\!\left(\!\frac{R\!-\!d_0}{d_0}\!\right)^{0.267}\right]\!
  \left\{\!1\!+\!\varepsilon \!\left[26.2\!-\!44\left(\!\frac{d_{0}}{R}\right)^{0.276}\! \right]\right\}
\end{equation}

\subsection{Discussion}

The analysis of axial stiffness of the ideal multiwalled nanotubes based on the van der Waals intershell interaction shows that only several MWCNT's external shells contribute to its total stiffness under the load imposed only on the external shell. The number of contributing shells in fact less than 5 for nanotubes with $R \lesssim 10$ nm and never exceed $15$ even for extremely thick nanotubes (figure \ref{RigidityMWCNT}). As a result the MWCNT actual stiffness may be several times lower than its ideal stiffness which is obtained under the assumption that all shells are loaded evenly. This fact should be taken into account while MWCNT-based towlines, cables and armoring elements are constructed.

Contrary to single-walled nanotubes, which are at least for small elongations subordinated to the Hooks law $F=k \varepsilon L,\;\;k=const$, the stiffness of MWCNTs even for rather weak loads depends linearly on the tube elongation ($k \sim \varepsilon$) as a result of the load-induced gradual involvement of inner shells.

All calculations presented here in support of our assertion should be performed for particular values of the Poisson ratio and sound velocity, which are (in the case of ideal lattice) uniquely determined only by the lattice structure and interatomic force constants. However up to now there is a large scatter of values of Poisson ratio obtained by different authors ($\nu=0.14-0.19$,\cite{PoissonRatioT} $\nu=0.17-0.22$,\cite{PoissonRatioG} $\nu=0.19$,\cite{Reinf3,Poisson2} $\nu=0.19-0.22$,\cite{Poisson3} $\nu=0.20-0.23$,\cite{Popov} $\nu=0.25 - 0.27$,\cite{Poisson1} $\nu=0.27 - 0.28$ \cite{Young7thMW}). In most cases $\nu\in 0.16 - 0.3$. Here we used the value $\nu=0.17$ which is a mean value of Poisson ratio for finite graphene sheets of different sizes \cite{PoissonRatioG} and chiral SWCNTs.\cite{PoissonRatioG} It is also close to the corresponding magnitude along the basal plane in graphite $\nu=0.16$.\cite{PoissonRatio1970}

In view of the existing data scattering we have looked how the calculated MWCNT stiffness depends on the used values of Poisson ratio and sound velocity. It clear that for Poisson ratios higher than $\nu=0.17$ the strain transmission from the outermost shell into the depth of the tube raises what leads to higher values of nanotube stiffness. However even for $\nu=0.3$ (which exceeds most of its theoretical and experimental estimates for SWCNTs) it appears that only $12-15$ external shells contribute to the $k$ even for MWCNTs with $R\geq 25$ nm and $\varepsilon\gtrsim0.05$. In other words, despite the fact that actual stiffness of MWCNT grows almost linearly with increasing of the shells' Poisson ratio moving to its "ideal" value, nevertheless the latter remains many times higher. ( The values of axial stiffness for varying values of Poisson ratio may be found with an accuracy of several per cents using the interpolating formula
$k(\nu)\approx k(\nu_{0}) \left(1+\frac{17}{20}\frac{\nu-\nu_{0}}{\nu_{0}}\right)$  ).

The sound velocity actually depends on the quality of the shell's lattice and connected with the stiffness of each shell and "ideal" total stiffness of MWCNT by the relation $k\sim c^{2}$. It appears that the sound velocity and the actual stiffness of MWCNT are linearly dependent in the interval $c=17000 - 23000\;km/sec$ as a consequence of a weakening of inner shells strains due to the preferred accumulation of the van der Waals interaction energy in subsurface shells. So if used here $c\!=\!18.4$ km/sec is underestimated the number of shells actually contributing to the MWCNT stiffness would be even lower than indicated above.

Note that the stiffness of perfect MWCNTs is perceptibly greater than that of MWCNTs with rather low defect concentration,\cite{YoungDef} but their ratios to the corresponding ideal values are in the opposite relation (i.e., the actual and ideal rigidities of some MWCNT would be most close in values in the case of pretty defect MWCNT).

It also should be noted, that some of MWCNT inner shells may be segmented. In accordance to our analysis such situations do not affect substantially the axial stiffness in all cases when the cumulative length of "broken" shell is close to length of its neighbors and gaps between shell's parts are about the range of van der Waals interaction ($\sim 1$ nm), that is the elastic properties of MWCNT are resistant to fracture and minor damage of some of its inner shells.

In that case when one of inner shells with number $i$ appears to be sufficiently shorter than neighboring shells the MWCNT can be considered as consistent of two placed in series parts with different number of walls and as a result different rigidities (the first part of length $L_{1}-L_{i}$ is a $(i-1)$-walled MWCNT and the second part of length $L_{i}$ is a $n$-walled MWCNT). The effective stiffness for such a series is $k=\frac{2k_{1}k_{2}}{k_{1}+k_{2}}$, where $k_{1}$ and $k_{2}$ are determined by (\ref{RigidityOfMWCNT}).

Thus for applications where the highest possible axial stiffness of single nanotube or nanotube bundle is needed the 4-5-walled nanotubes of minimal diameter are quite sufficient.

\subsection*{Acknowledgements}
Author is grateful to Prof. Vadim Adamyan for discussions and remarks. This work was supported by the Ministry of Education and Science of Ukraine, Grant \#0109U000929.

% ----------------------------------------------------------------------------------------------------------------------------------------------------


\section*{References}
\begin{thebibliography}{35}

\bibitem{Reinf1} R. S. Ruoff and D. C. Lorents, Carbon \textbf{33}, 925 (1995).
\bibitem{Reinf2} S. Govindjee and J. L. Sackman, Solid State Commun. \textbf{110}, 227 (1999).
\bibitem{Reinf3} B. I. Yakobson and Ph. Avouris, in \textit{Mechanical Properties of Carbon Nanotubes}, edited by M. S. Dresselhaus, G. Dresselhaus, Ph. Avouris, (Springer–Verlag, Heidelberg, Germany, 2001), Topics in Applied Physics \textbf{80}, 287.
\bibitem{fracture1} B. I. Yakobson, C. J. Brabec, and J. Bernholc, Phys. Rev. Lett. \textbf{76}, 2511 (1996).
\bibitem{fracture2} M. F. Yu, O. Lourie, M. J. Dyer, K. Moloni, T. F. Kelly, and R. S. Ruoff, Science \textbf{287}, 637 (2000).
\bibitem{Revers1} S. Iijima, C. Brabec, A. Maiti, and J. Bernholc, J. Chem. Phys. \textbf{104}, 2089 (1996).
\bibitem{Revers2} D. A. Walters, L. M. Ericson, M. J. Casavant, J. Liu, D. T. Colbert, K. A. Smith, and R. E. Smalley, Appl. Phys. Lett. \textbf{74}, 3803 (1999).
\bibitem{Revers3} M. F. Yu, B. S. Files, S. Arepalli, and R. S. Ruoff, Phys. Rev. Lett. \textbf{84}, 5552 (2000).
\bibitem{Revers4} T. W. Tombler, C. Zhou, J. Kong, H. Dai, L. Liu, C. S. Jayanthi, M. Tang, and S. Y. Wu, Nature \textbf{405}, 769 (2000).
\bibitem{Young1} M. M. J. Treacy, T. W. Ebbesen, and J. M. Gibson, Nature \textbf{381}, 678 (1996).
\bibitem{Young2} A. Krishnan, E. Dujardin, T. W. Ebbesen, P. N. Yianilos, and M.M.J. Treacy, Phys. Rev. B \textbf{58}, 14013 (1998).
\bibitem{Young3} E. W. Wong, P. E. Sheehan, and C. M. Lieber, Science \textbf{277}, 1971 (1997).
\bibitem{Young4} J. P. Salvetat, G. A. D. Briggs, J. M. Bonard, R.R. Bacsa, A. J. Kulik, T. St\"{o}ckli, N. A. Burnham, and L. Forr\'{o}, Phys. Rev. Lett. \textbf{82}, 944 (1999).
\bibitem{Young5}  B. G. Demczyk, Y. M. Wang, J. Cumings, M. Hetman, W. Han, A. Zettl, and R. O. Ritchie, Mater. Sci. Eng., A \textbf{334}, 173 (2002).
\bibitem{Young6MW} Z. W. Pan, S. S. Xie, L. Lu, B. H. Chang, L. F. Sun, W. Y. Zhou, G. Wang, and D. L. Zhang, Appl. Phys. Lett. \textbf{74}, 3152 (1999).
\bibitem{Young8thCont} P. Zhang, Y. Huang, P. H. Geubelle, P. A. Klein, and K. C. Hwang, J. Solids Struct. \textbf{39}, 3893 (2002).
\bibitem{Young9} Y. Wu, M. Huang, F. Wang, X. M. Henry Huang, S. Rosenblatt, L. Huang, H. Yan, S. P. O'Brien, J. Hone, and T. F. Heinz, Nano Lett. \textbf{8}, 4158 (2008).
\bibitem{MWCNTDepOnRadius1} Z. Tu and Z. Ou-Yang, Phys.Rev. B \textbf{65}, 233407 (2002).
\bibitem{Young7thMW} J. P. Lu, Phys. Rev. Lett. \textbf{79}, 1297 (1997).
\bibitem{InverseRadiusPoncharal} P. Poncharal, Z. L. Wang, D. Ugarte, and W. A. de Heer, Science \textbf{283}, 1513 (1999).
\bibitem{MWCNTDepOnRadius2} N. Yao and V. Lordi, J. Appl. Phys. \textbf{84}, 1939 (1998).
\bibitem{SWCNTNotDepOnRadius} Z. Xin, Z. Jianjun, and O.-Y. Zhong-can, Phys. Rev. B \textbf{62}, 13692 (2000).
\bibitem{MolecDyn1} J.-Y. Hsieh, J.-M. Lu, M.-Y. Huang, and C.-C. Hwang, Nanotechnology \textbf{17}, 3920 (2006).
\bibitem{MolecDyn2} Z. Peralta-Inga, S. Boyd, J. S. Murray, C. J. O'Connor, and P. Politzer, Struct. Chem. \textbf{14}, 431 (2003).
\bibitem{Adamyan} V. Adamyan and V. Zavalniuk, J.Phys.: Condens. Matter \textbf{23}, 015402 (2010).
\bibitem{Marchenko} V. Zavalniuk and S. Marchenko, Fiz. Nizk. Temp. \textbf{37}, 432 (2011).
\bibitem{girifalco} L. A. Girifalco, M. Hodak, and R. S. Lee, Phys. Rev. B \textbf{62}, 013104 (2000).
\bibitem{PoissonRatio1970} O. L. Blakslee, D. G. Proctor, E. J. Seldin, G. B. Spence, and T. Weng, J. Appl. Phys. \textbf{41}, 3373 (1970).
\bibitem{PoissonRatioT} D. S\'{a}nchez-Portal, E. Artacho, J.M. Soler, A. Rubio, and P. Ordej\'{o}n, Phys.Rev. B \textbf{59}, 12678 (1999).
\bibitem{PoissonRatioG} J.-W. Jiang, J.-S. Wang, and B. Li, Phys.Rev. B \textbf{80}, 113405 (2009).
\bibitem{Poisson2} B. I. Yakobson, C. J. Brabec, and J. Bernholc, Phys. Rev. Lett. \textbf{76}, 2511 (1996).
\bibitem{Poisson3} A. Sears and R. C. Batra, Phys. Rev. B \textbf{69}, 235406 (2004).
\bibitem{Popov} V. N. Popov, V. E. Van Doren, and M. Balkanski, Phys. Rev. B \textbf{61}, 3078 (2000).
\bibitem{Poisson1} Y. Jin and F. G. Yuan, Compos. Sci. Technol. \textbf{63}, 1507 (2003).
\bibitem{YoungDef} J. P. Salvetat, A. J. Kulik, J. M. Bonard, G. A. D. Briggs, T. St\"{o}ckli, K. M\'{e}t\'{e}nier, S. Bonnamy, F. B\'{e}guin, N. A. Burnham, and L. Forr\'{o}, Adv. Mater. \textbf{11}, 161 (1999).

\end{thebibliography}
\end{document}